# Realization of Pristine and Locally-Tunable One-Dimensional Electron Systems in Carbon Nanotubes

J. Waissman[1,2]*, M. Honig[1]*, S. Pecker[1]*, A. Benyamini[1]*, A. Hamo[1]* and S. Ilani[1]

[1]*Department of Condensed Matter Physics, Weizmann Institute of Science, Rehovot 76100, Israel*

[2]*School of Engineering and Applied Sciences, Harvard University, Cambridge, Massachusetts 02138, USA*

**Recent years have seen the development of several experimental systems capable of tuning local parameters of quantum Hamiltonians. Examples include ultracold atoms[1], trapped ions[2], superconducting circuits[3], and photonic crystals[4]. By design, these systems possess negligible disorder, granting them a high level of tunability. Conversely, electrons in conventional condensed matter systems exist inside an imperfect host material, subjecting them to uncontrollable, random disorder, which often destroys delicate correlated phases and precludes local tunability. The realization of a condensed matter system that is disorder-free and locally-tunable thus remains an outstanding challenge. Here, we demonstrate a new technique for deterministic creation of locally-tunable, ultra-low-disorder electron systems in carbon nanotubes suspended over circuits of unprecedented complexity. Using transport experiments we show that electrons can be localized at any position along the nanotube and that the confinement potential can be smoothly moved from location to location. Nearly perfect mirror symmetry of transport characteristics about the centre of the nanotube establishes the negligible effects of electronic disorder, thus allowing experiments in precision engineered one-dimensional potentials. We further demonstrate the ability to position multiple nanotubes at chosen separations, generalizing these devices to coupled one-dimensional systems. These new capabilities open the door to a broad spectrum of new experiments on electronics, mechanics, and spins in one dimension.**

The carbon nanotube is a promising substrate for realizing an ultra-clean and locally-tunable electron system. Contrary to conventional semiconductors, carbon nanotubes have been shown to naturally grow exceptionally cleanly, leading to low inherent disorder[5]. Moreover, the long lengths of this one-dimensional (1D) system portend the possibility to control the potential at each point along its length using an array of transverse electrostatic gates (Fig.1a). Nanotubes also possess a collection of desirable physical properties[6]: Their strong electron-electron interactions could generate correlated electronic ground states[7–10], the ability to localize and control individual spins could realize a quantum information chain or charge/spin pumps[11–17], and these can interact with its mechanical motion[18–21] or with correlated materials[22–26].

To date, studies have exploited these properties mostly in zero-dimensional single and double quantum dot settings. The extension to longer 1D settings has so far been hindered by disorder, which at low temperatures breaks the electron system into localized, uncontrolled quantum dots[27,28]. The bottleneck is in conventional technologies for making ultra-clean nanotube devices, which require two demanding processes to succeed simultaneously: the growth of pristine nanotubes and the fabrication of complex electrical circuits. Nanotube cleanliness is achieved by growing the nanotubes as the last step in device fabrication[5], which limits device design due to the high temperature of nanotube growth. Recent stamping approaches[17,29] have eliminated some of these issues by growing nanotubes separately from the measurement circuit and transferring them mechanically. However, these approaches remain statistical in nature, resulting in a small yield of a few percent even for simple and short devices. Increasing the device complexity with either longer nanotubes or more complex circuits will decrease the yield further, rendering these approaches less practical. Thus, the potential of the nanotube as a system for locally-tunable experiments in extended 1D geometries remains unrealized.

In this paper we report the realization of a new nano-assembly technique that allows us to deterministically create ultra-low disorder, suspended, multi-nanotube devices with electrical circuits of arbitrary complexity. Our approach uses scanning probe microscope manipulation to achieve deterministic assembly. On one chip we grow long, parallel nanotubes suspended without slack over wide trenches (Fig.1b; see Methods). On a separate chip we fabricate the electrical circuit on a narrow cantilever (Fig.1c). For the device in Fig.1a, for example, an array of parallel electrodes is fabricated, where the external ones (contacts, yellow) are taller than the rest (gates, blue). The nanotubes and circuits are fabricated in two independent processes and neither one imposes any restrictions on the other. The scanning probe microscope is then used to insert the cantilever into a trench and "mate" it with a nanotube (Fig.1d). Since the contacts are taller than the gates, the nanotube touches them first and remains suspended over the gates (Fig.1e). Importantly, this process works at low temperatures ($T = 4K$), where electronic cleanliness can be tested via *in-situ* transport measurements. By mating to nanotubes in different trenches, or to different segments of one nanotube, we can therefore select only perfectly clean tubes, with chosen bandgaps, and assemble them into complex devices in a deterministic way.

To find the desired nanotube we must browse through several trenches, connecting and detaching from different nanotubes without crashing or contaminating the cantilever. To do this safely without visual aids we implemented a capacitance-based detection scheme (Methods) that together with piezoelectric positioners allows control of their relative position with $\sim 1\mu m$ accuracy. The final approach is done by a piezoelectric scanner with nanometre accuracy, allowing the cantilever to gently touch the tube at any depth inside the trench, and to finely control its lateral position along the circuit. Contact to a nanotube is identified by a change in resistance between the two chips, and the circuit is lowered until the nanotube touches all contact electrodes. At this point, we measure *in-situ* gate-dependent transport. Having identified a desirable nanotube, we pass high current through adjacent pairs of contacts at the sides of the device to surgically cut the nanotube at well-defined locations and separate it from the nanotube chip without damaging the segment above the gates (Fig.1e and Supplementary S1).

Figure 2c shows a representative and unprecedented seven-gate device made by our mating technique using nanotube and circuit chips similar to those shown in figures 2a and b. The nanotube is perpendicular to the gates, suspended without slack over a length of 1.2µm at a fixed height of 130nm above all gates, anchored over the entire length of the contacts and does not touch silicon oxide, characteristics achieved in the vast majority of mated devices (Supplementary S2). To localize multiple nanotubes at specific locations, we pattern contacts of different lengths and move the cantilever along the trench with the piezo-scanner, successively touching and detaching from a tube, until resistance measurements indicate that it is touching contacts corresponding to a specific location. Figure 2d shows a double-nanotube device made with this technique. The first nanotube is positioned on a set of shorter contacts, with matching gates, and cut to electrically isolate it from all other contacts. The second nanotube is then positioned on a longer set of contacts, with a second set of gates wrapped around the longer contacts, allowing both nanotubes to be independently contacted and gated (Supplementary S8 shows measurements demonstrating electrostatic coupling of the two nanotubes). Geometrically, the device in figure 2c is the closest yet achieved to the ideal illustrated in Fig.1a, and the device in figure 2d extends this capability to a new class of devices with multi-nanotube geometry.

Are these devices as ideal electronically as they are geometrically? We address this question using transport measurements of multiply-gated devices with small bandgap nanotubes. We start with the simplest experiment on a five-gated device, with all gates chained together, reproducing past single-gate transistor experiments. We use gold contacts that dope the nanotube segments above them with holes, and control the doping of the suspended nanotube segment electrostatically with the gates. As a function of the gate voltage the conductance measured at $T = 4$K shows three regimes (Figure 3a): At negative voltages the suspended segment is hole-doped, forming a continuous "nanotube wire" whose conductance is weakly gate-dependent[27]. At intermediate voltages the nanotube is doped into its bandgap, determined to be $34 \pm 5$meV from finite bias measurements, and the conductance is suppressed. For positive voltages the suspended segment is doped with electrons, forming a pair of *p-n* junctions near the contacts which confine a large quantum dot, whose charging by individual electrons generates Coulomb blockade oscillations in the conductance. The oscillation periodicity, $\Delta V_g = 31.5 \pm 1.5$mV, given by $\Delta V_g = e/C_g$ with $e$ the electron charge and $C_g$ the gate capacitance, agrees well with that expected from the capacitance of the length of the suspended segment, $L = 880$nm, to all five gates. The corresponding charging energy of this large dot, obtained from finite-bias Coulomb diamonds, is $E_C = 10 \pm 2$meV. The clean and regular spectrum of oscillations therefore signifies the formation of a single quantum dot over the entire suspended nanotube length, whose electronic cleanliness is comparable to the best ultra-clean nanotube devices made to date[5,9,12,30].

The local gates now allow us to probe electronic behavior on finer spatial scales. By electron-doping the nanotube locally with a single gate and hole-doping the rest of it with all other gates, we form a smaller quantum dot localized above this gate. Accordingly, with five independent gates we can form, in principle, dots at five different locations along the nanotube, whose characteristics reflect the spatial dependence of the nanotube electronic properties. Figure 3b shows the corresponding five conductance traces, as a function of the individual gate voltages. Clearly, single quantum dots are formed at all positions. Their Coulomb blockade oscillations have periodicities of $\Delta V_g = 280 \pm 10$mV, indicating that the dots are well localized above a single gate (Supplementary S4). The corresponding charging energy of these dots is $E_C = 59 \pm 8$meV. Moreover, all traces exhibit a single periodicity, showing that the dots are clean. The Coulomb peak heights, however, vary between dots at different locations, hinting at possible position-dependence in the electronic properties along the nanotube.

A more complete picture of the spatial dependence is obtained by using pairs of gates to continuously move a quantum dot along the nanotube. Figure 3c shows the conductance measured as a function of the voltages on gates 1 and 2, while all other gates are negatively biased. On the bottom left corner, both gates dope the nanotube with holes and no dot is formed. When gate 1 (gate 2) is positively biased, along the horizontal (vertical) axis, a dot forms above this gate. Biasing both gates together (upper-right corner) extends the dot above both gates. Thus, going from the bottom-right to the top-left of this figure the quantum dot shifts from one gate to its neighbor. In this measurement, the Coulomb charging peaks appear as charging lines, separating different charge states of the quantum dot. Their local slope corresponds to the relative capacitance of the dot to the two gates, and reflects the position of the centre-of-mass of the electronic charge. Notably, the slopes of all charging lines, down to that of the first electron, evolve smoothly and monotonically during the shift, reflecting the smooth transfer of the electronic confinement from site to site. The data, however, contain unexpected features: the charging lines exhibit periodic stripe modulation of the peak heights and a band-like region where the conductance is suppressed (arrows, Fig. 3c). These features may indicate the existence of disorder forming random barriers or dots. Below we show, however, that these features arise from intrinsic electrostatics, and not disorder.

A clear way to determine whether the observed features are due to disorder is to perform the mirror-symmetric version of the experiment depicted in Fig. 3c. This measurement, shown in Fig. 3d, is done with the opposite gates, 4 and 5, over a voltage range identical to that in Fig 3c. Comparing these mirror-symmetric measurements reveals a striking similarity: charging line slopes, positions, and spacing are all identical. Furthermore, the peak modulations and the conductance suppression are reproduced at the same gate voltages. The remarkable implication is that all the observed features are not the result of a random disorder potential, but rather arise from the intrinsic electrostatics of the device. These features, discussed further in Supplementary S5, are due to gating of nanotube segments that are beyond the dot, such as Fabry-Perot-like oscillations in the hole-doped "nanotube leads". In contrast to peak positions, which are identical in both experiments, peak heights are different. While peak positions are sensitive only to electrostatics, their heights also depend on the resistance to the metal contacts, which might vary for different contacts. We find, however, that this asymmetry does not originate in contact resistance but instead is also electrostatic in nature, coming from an inequivalence of the *p-n* junction barriers near the source and drain contacts, due to a slight lithographic misalignment (~15nm) of the gates toward the drain contact. The observation of nearly-perfect mirror symmetry thus demonstrates that, for electrons above the outer gates, electrostatics rather than random disorder determines the local electronic structure.

To check the effects of disorder in the bulk of the suspended nanotube, we generalize the above measurements to all pairs of gates in the device. Figure 4 shows a matrix of two-gate conductance measurements, whose columns and rows correspond to the gates scanned on the horizontal and vertical axes of each panel. In all panels the gate voltage ranges are identical, with all other gates maintaining a constant hole-doping voltage. On the main diagonal of this matrix, the scanned gates are nearest neighbors (as in Fig. 3c,d). Clearly, all the scans along this diagonal feature a continuous bending of the charging lines, indicating the smooth movement of charge from any gate to its neighbor. Scans with non-adjacent gates form two or more quantum dots along the nanotube. While many features are observed in these experiments, the remarkable observation is that over the entire matrix all these features are symmetric among experiments with mirror symmetry around the nanotube centre (dashed black line). We conclude that, to the spatial resolution fixed by our gates and to the energy scale set by the temperature, disorder is playing a negligible role in determining the potential landscape along the entire device. Measurements on a different device show similar device cleanliness down to dilution refrigerator temperatures (Supplementary S7).

The last step to establish our system as a controllable laboratory for 1D experiments is to demonstrate a quantitative understanding of its electrostatics, permitting 1D potential design. A potential can be straightforwardly defined if the gates are placed close to the suspended nanotube compared to their pitch. Close gates, however, screen electron-electron interactions, destroying this highly-desirable feature. It is therefore beneficial to distance the gates from the nanotube, but this results in non-local gating, and potential design then requires a quantitative accounting for this non-locality. This information, how a gate at position $i$ influences the nanotube at a position above gate $j$, is embedded in figure 4: At the bottom right corner of each $(i,j)$ panel a quantum dot is localized over gate $i$ and gated by gate $j$. The dot, acting as a local charge detector, allows us to directly measure the capacitive coupling elements $C_{ij}$ (25 in total, Supplementary S3 for details). In the inset of figure 4 we present the extracted capacitance distribution for each gate (points), and compare it with calculated capacitance distributions from a finite-element simulation (lines). These calculations, with no free parameters, fit the experimental points quantitatively well. Specifically, we see that the capacitance distributions of gates 2-4 are almost identical, demonstrating that the electrostatics in the "bulk" of the sample is translationally invariant. The edge gates (1 and 5) show reduced coupling due to screening by the contacts, and differ in their peak coupling due to the gate-contact misalignment asymmetry noted above, fully reproduced by the calculations.

These results are not sensitive to mechanical displacement of the nanotube, which for this device, with 1V applied to all five gates, is estimated to be ~5nm, small compared to the 130nm gate-nanotube distance. Using these calculations, we put an upper bound of $\sim 5 \cdot 10^{-2} e$ on the level of charge disorder on 100nm length scales, corresponding to ~5mV of bare disorder potential or ~50μ$V$ in the self-consistent disorder potential (Supplementary S4). These low values of disorder and the correspondence of the measured and calculated electrostatics show that potential profiles can be accurately designed. In Supplementary section S6, we show how detailed knowledge of the capacitive coupling elements $C_{ij}$ allows us to deconvolve the effect of the nanotube-gate separation and design potentials with resolution limited only by the density of gates.

The ability to identify perfect nanotubes and selectively nano-assemble them at predefined positions in an electronic circuit makes possible devices that were previously inconceivable. Currently, we assemble electronically-pristine 1-2μm-long multi-gated devices in a span of a few hours, suggesting that even far more complex devices are possible with this technique. These new devices constitute a novel laboratory for studying electronic phases of strongly-interacting electrons in 1D, subject to engineered potentials. They also act as clean mechanical resonators that can now be coupled to multiple quantum dots. Furthermore, we demonstrate a new class of devices involving multiple NTs positioned at chosen locations, heralding sensitive local charge detectors and coupled 1D systems (see Supplementary S8). We expect these novel devices to lead to a new wave of experiments in nanotubes with applications ranging from fundamental condensed matter physics to nano-electromechanics and quantum information science.

**Methods**

The "nanotube chip" is formed in a two-step etching process. The first etch, in KOH solution, forms deep, wide trenches with a ~50° wall angle, while the second etch, in TMAH solution, forms a shallow trench lip with a ~23° angle (see Fig.2a, inset). This shallow lip allows nanotubes to easily stick to the surface after growth, which eliminates their slack. The chip is then metallized with Ti/Pt (5nm/150nm, respectively). Nanotubes are grown from catalyst deposited on the plateaus between trenches in lithographically-defined pads. The growth is performed with Chemical Vapor Deposition using a standard growth recipe for single-walled carbon nanotubes, with argon, hydrogen, and ethylene gases. Feedstock gas flow alignment ensures growth of parallel suspended nanotubes ($\pm 3°$).

The "circuit chip" is patterned on a Si/SiO$_2$ wafer using e-beam lithography, followed by the evaporation of contacts (5nm/150nm Ti/Au), gates (5nm/20nm Ti/PdAu) and deep reactive ion etching with lithographically-defined etch masks.

We found that an important step to establish good electrical and mechanical contact between a nanotube and contact electrodes is *in-situ* cleaning of the contact surfaces using argon ion etching in a load-lock. Immediately after this etching the sample is inserted into the microscope for mating. With this step, we achieve mechanically-stable contacts with typical resistances of ~100kΩ measured at room temperature. These values are typical for these kinds of devices, although still falling short of the resistances demonstrated[31] in some devices that approach $R = h/4e^2$.

Blind navigation in the microscope is performed with capacitance measurements. Four metal pads are patterned on the measurement circuit chip, two large pads of area ~75,000µm² each and two small pads of ~5,000µm² each. Their capacitance is measured with respect to the metallized nanotube chip, which serves as the second electrode. The pads are designed such that scanning allows us to establish, with rough and fine resolution, the relative position and angle of the two chips. The capacitance is measured using displacement current at a frequency of ~12kHz, as a function of nano-positioner movement (Attocube). First, the separation between the two chips and the orientation and relative position of their edges is determined. A final scan with a small capacitor parallel to the trenches establishes the trench positions for mating. Overall, we determine the relative position and orientation of the two chips in all three dimensions to within $\sim 1\mu m$ and $\sim 0.1°$.

Conductance measurements are performed with standard AC lock-in techniques, using a 100 µV excitation.

**References**

1. Bloch, I., Dalibard, J. & Nascimbène, S. Quantum simulations with ultracold quantum gases. *Nature Physics* **8**, 267–276 (2012).

2. Blatt, R. & Roos, C. F. Quantum simulations with trapped ions. *Nature Physics* **8**, 277–284 (2012).

3. Houck, A. a., Türeci, H. E. & Koch, J. On-chip quantum simulation with superconducting circuits. *Nature Physics* **8**, 292–299 (2012).

4. Aspuru-Guzik, A. & Walther, P. Photonic quantum simulators. *Nature Physics* **8**, 285–291 (2012).

5. Cao, J., Wang, Q. & Dai, H. Electron transport in very clean, as-grown suspended carbon nanotubes. *Nature materials* **4**, 745–9 (2005).

6. Jorio, Ado, Dresselhaus, Gene, Dresselhaus, Mildred S. (Eds.) *Carbon Nanotubes: Advanced Topics in the Synthesis, Structure, Properties and Applications*. (Springer: 2008).

7. Yao, Z., Postma, H. W. C., Balents, L. & Dekker, C. Carbon nanotube intramolecular junctions. *Nature* **402**, 273–276 (1999).

8. Bockrath, M. *et al.* Luttinger-liquid behaviour in carbon nanotubes. *Nature* **397**, 598–601 (1999).

9. Deshpande, V. V. & Bockrath, M. The one-dimensional Wigner crystal in carbon nanotubes. *Nature Physics* **4**, 314–318 (2008).

10. Deshpande, V. V *et al.* Mott insulating state in ultraclean carbon nanotubes. *Science* **323**, 106–10 (2009).

11. Loss, D. & DiVincenzo, D. P. Quantum computation with quantum dots. *Physical Review A* **57**, 120–126 (1998).

12. Kuemmeth, F., Ilani, S., Ralph, D. C. & McEuen, P. L. Coupling of spin and orbital motion of electrons in carbon nanotubes. *Nature* **452**, 448–52 (2008).

13. Buitelaar, M. *et al.* Adiabatic Charge Pumping in Carbon Nanotube Quantum Dots. *Physical Review Letters* **101**, 126803 (2008).

14. Churchill, H. *et al.* Relaxation and Dephasing in a Two-Electron $^{13}$C Nanotube Double Quantum Dot. *Physical Review Letters* **102**, 2–5 (2009).

15. Kuemmeth, F., Churchill, H. O. H., Herring, P. K. & Marcus, C. M. Carbon nanotubes for coherent spintronics. *Materials Today* **13**, 18–26 (2010).

16. Jespersen, T. S. *et al.* Gate-dependent spin–orbit coupling in multielectron carbon nanotubes. *Nature Physics* **7**, 348–353 (2011).

17. Pei, F., Laird, E. A., Steele, G. A. & Kouwenhoven, L. P. Valley–spin blockade and spin resonance in carbon nanotubes. *Nature Nanotechnology* **7,** 630-634 **(2012).**

18. Sazonova, V. *et al.* A tunable carbon nanotube electromechanical oscillator. *Nature* **431**, 284–7 (2004).

19. Leturcq, R. *et al.* Franck–Condon blockade in suspended carbon nanotube quantum dots. *Nature Physics* **5**, 327–331 (2009).

20. Steele, G. A. *et al.* Strong coupling between single-electron tunneling and nanomechanical motion. *Science* **325**, 1103–7 (2009).

21. Lassagne, B. *et al.* Coupling mechanics to charge transport in carbon nanotube mechanical resonators. *Science* **325**, 1107–10 (2009).

22. Sahoo, S. *et al.* Electric field control of spin transport. *Nature Physics* **1**, 99–102 (2005).

23. Jarillo-Herrero, P., Van Dam, J. A. & Kouwenhoven, L. P. Quantum supercurrent transistors in carbon nanotubes. *Nature* **439**, 953–6 (2006).

24. Hauptmann, J. R., Paaske, J. & Lindelof, P. E. Electric-field-controlled spin reversal in a quantum dot with ferromagnetic contacts. *Nature Physics* **4**, 373–376 (2008).

25. Pillet, J.-D. *et al.* Andreev bound states in supercurrent-carrying carbon nanotubes revealed. *Nature Physics* **6**, 965–969 (2010).

26. Schindele, J., Baumgartner, A. & Schönenberger, C. Near-Unity Cooper Pair Splitting Efficiency. *Physical Review Letters* **109**, 157002– (2012).

27. McEuen, P., Bockrath, M., Cobden, D., Yoon, Y.-G. & Louie, S. Disorder, Pseudospins, and Backscattering in Carbon Nanotubes. *Physical Review Letters* **83**, 5098–5101 (1999).

28. Woodside, M. T. & McEuen, P. L. Scanned probe imaging of single-electron charge states in nanotube quantum dots. *Science* **296**, 1098–101 (2002).

29. Wu, C. C., Liu, C. H. & Zhong, Z. One-step direct transfer of pristine single-walled carbon nanotubes for functional nanoelectronics. *Nano Letters* **10**, 1032–6 (2010).

30. Steele, G. A., Gotz, G. & Kouwenhoven, L. P. Tunable few-electron double quantum dots and Klein tunnelling in ultraclean carbon nanotubes. *Nature Nanotechnology* **4**, 363–7 (2009).

31. Liang, W. *et al.* Fabry - Perot interference in a nanotube electron waveguide. *Nature* **411**, 665–9 (2001).

**\*** These authors contributed equally to this work

**Acknowledgements:** We acknowledge N. Shadmi and E. Joselevich for nanotube growth at initial stages of the project, D. Mahalu for the e-beam writing, A. Yoffe and S. Garusi for the dry etching and F. Kuemmeth, P. McEuen, H. Shtrikman, F. von-Oppen and A. Yacoby for comments on the manuscript. S.I. acknowledges the financial support by the ISF Legacy Heritage foundation, the Bi-National Science Foundation (BSF), the Minerva Foundation, the ERC Starters grant, the Marie Curie People grant (IRG), and the Alon Fellowship. S.I. is incumbent of the William Z. and Eda Bess Novick career development chair.

**Author Contributions:** All authors conceived, designed and performed the experiments. JW, SP, AB, AH and SI analyzed the data. SP, AB and AH contributed analysis tools. JW and SI wrote the paper.

**Additional Information:** Supplementary information accompanies this paper at www.nature.com/naturenanotechnology. Reprints and permission information is available online at http://npg.nature.com/reprintsandpermissions. Correspondence and requests for materials should be addressed to SI.

**Figure 1: Illustration of the nano-assembly technique for creating clean and complex nanotube devices.** a) An example of a device with desirable characteristics: a nanotube connected to source and drain electrodes (yellow) and suspended above multiple gates (blue). We assemble such a device from two independent chips: b) The "nanotube chip" with parallel nanotubes grown over wide trenches. c) The "circuit chip" consisting of contact electrodes (yellow) and gate electrodes (blue) formed on a narrow cantilever. Typical dimensions are indicated. d) The nano-assembly is achieved with a scanning probe microscope (illustrated), which controls the relative position of the two chips with high precision (arrows indicate directions of motion). e) A device is made by inserting the cantilever into a trench and "mating" the electrical circuit to several nanotubes until a desirable one is found. The nanotube touches the taller metallic contacts and remains suspended over the gates, allowing *in-situ* transport measurements (inset). Once a desirable nanotube is identified, it is locally cut by passing a large current between adjacent pairs of side contacts, without damaging the suspended segment (Supplementary S1), disconnecting the device from the nanotube chip.

**Figure 2: Individual components of the "mating" technique and representative nano-assembled devices.** a) Top: scanning electron microscope (SEM) picture of the "nanotube chip", comprising 30 trenches, ~100 μm-wide, etched in a silicon wafer and metalized with platinum. Suspended nanotubes are grown from the plateaus between trenches (Methods) (scale bar 50μm). Bottom: Zoom-in on a trench edge showing a single nanotube growing across the trench. The nanotube growth direction is aligned perpendicular to the trenches by the feedstock gas flow. A shallow slope at the trench edge (Methods) allows the nanotubes to easily stick to the surface, removing their slack (angles of the two slopes are indicated; scale bar 4μm). b) SEM picture of the "circuit chip", patterned on a Si/SiO$_2$ wafer, with 155nm high gold contacts and 25nm high PdAu gates. Deep etching leaves these electrodes on a thin (~10μm) and tall (~100μm) cantilever (scale bar 20μm). Inset: zoom-in to the cantilever tip (scale bar 2μm) c) A nano-assembled device with a single nanotube connected to contacts (yellow) and suspended at a height of 130nm over seven gates (blue, 150nm pitch) (scale bar 200nm). d) A two-nanotube device: The first nanotube sits on three contacts with two matching gates. The second nanotube is suspended over five gates, which wrap around its contacts

for independent addressability. Here, the shortest nanotube-nanotube distance is ~300nm, and the accuracy of positioning the nanotube from the opposite contact edge is less than 85nm. After mating, the nanotubes were selectively cut at two adjacent contact pairs (visible e.g. for nanotube 1; Supplementary S1) isolating the two devices from each other (scale bar 300nm) (see Supplementary S8 for measurements of this device).

**Figure 3: Localizing and moving electrons in clean quantum dots on a five-gated, small-bandgap nanotube device.** a) Top: SEM image of a device similar to the one measured (gate numbers are indicated). The nanotube is locally coloured according to its doping: holes – blue, electrons – red. The suspended segment is electrostatically doped by the gates while the segments above the contacts are hole-doped by the metal. Bottom: conductance, $G$, measured as a function of a common voltage on all five gates, $V_g$. Coulomb oscillations are apparent at positive gate voltages due to the formation of a quantum dot extended over the entire suspended nanotube. Insets show position-dependent nanotube band diagrams in the three different conductance regimes: a hole-doped "nanotube wire", the nanotube bandgap, and electron Coulomb oscillations (hole band – blue, electron band – red). b) Similar measurements as a function of voltages on five individual gates, $V_{gi}$ ($i$ the gate index), while the other gates maintain fixed hole-doping voltage, $V_{gj} = -0.8\text{V}, j \neq i$. In each trace a small electron quantum dot is formed above the corresponding gate (side illustrations) c) Conductance, $G$ (colourmap), measured as a function of the voltages on two adjacent gates, $V_{g1}$ and $V_{g2}$ (top illustration). Corner overlays show schematic band diagrams for different applied voltages. From the bottom-right to the top-left corner a dot is continuously shifted between the two adjacent gates. d) Similar measurement for the mirror-symmetric experiment with gates 4 and 5. While the values of conductance differ between panels c and d due to different *p-n* junction barriers formed near the left and right contacts, the conductance patterns are remarkably similar, down to small details (see main text).

**Figure 4: Characterization of the disorder and local electrostatic environment of the nanotube.** Main panel: A matrix of conductance measurements where in each entry a different pair of gate voltages, $V_{gi}$ and $V_{gj}$, is scanned. The gate scanned along the horizontal (vertical) axis is indicated in the column (row) title, and the voltage scan range

in all panels is identical to that in Figs. 3c,d. The detailed conductance features in all panels show symmetry with respect to mirror reflection around the nanotube centre (dashed black line). Colourmaps for all measurements are shown in the lower left corner of each entry, over the range 0…80nS for all scans. Inset: Electrostatic coupling between individual gates and the nanotube. Same-colour points show the extracted capacitive coupling between a given gate and five different quantum dots formed along the nanotube. Corresponding lines show the capacitance distribution of the gate to the nanotube calculated with a finite-element simulation incorporating the full device geometry, including the gate-contact misalignment of ~15nm (see Supplementary S3). Without any free parameters, the two show excellent agreement.

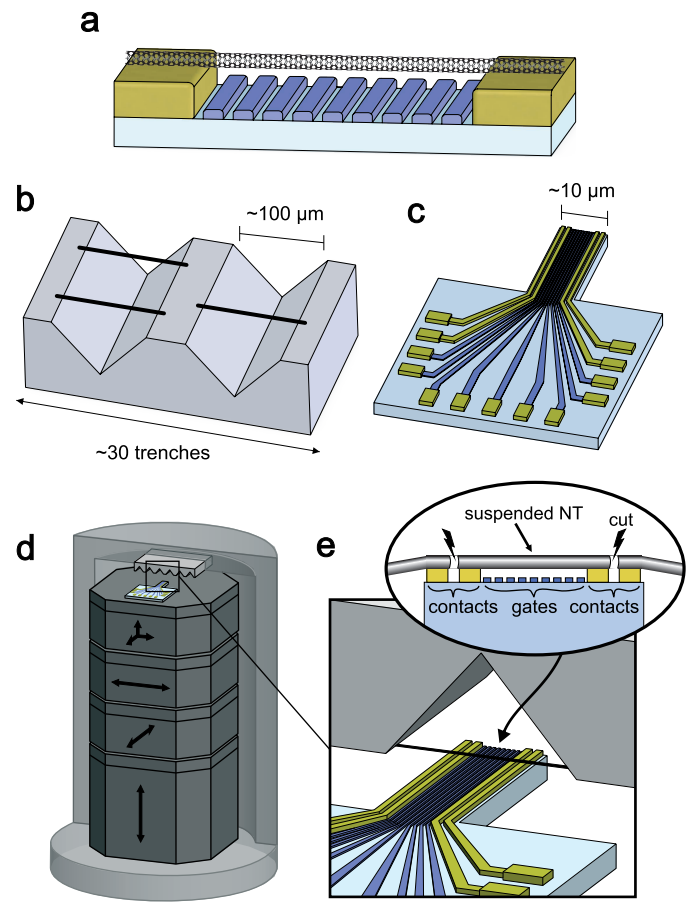

Figure 1

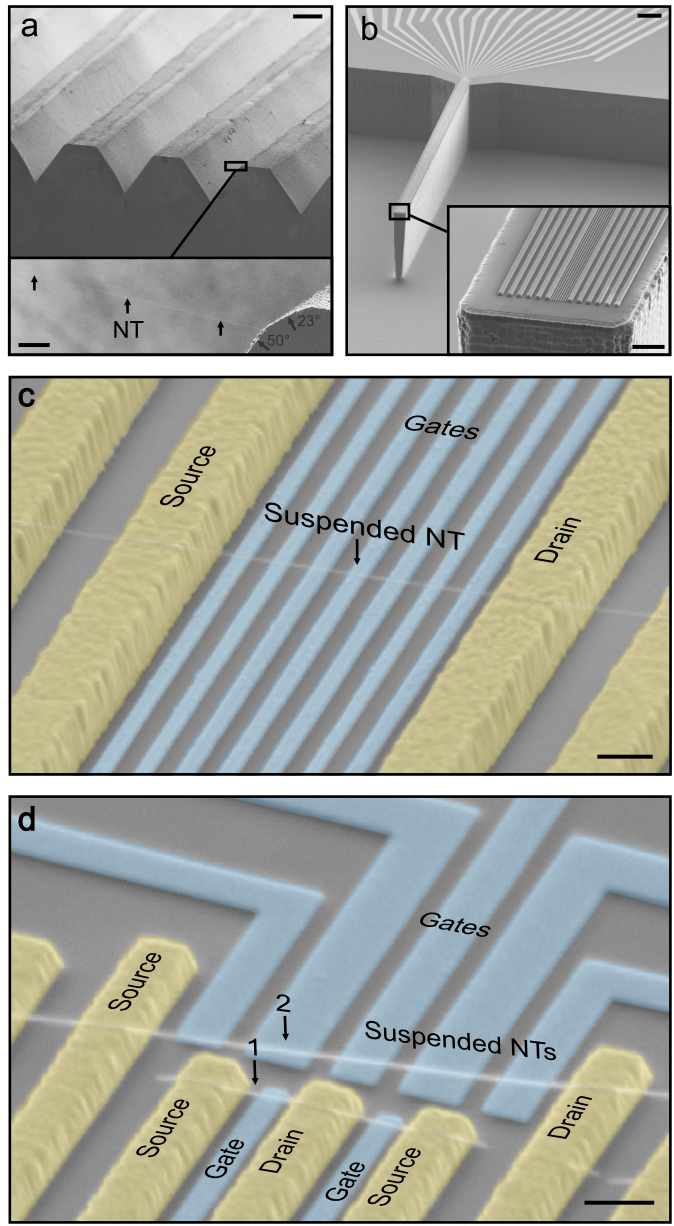

Figure 2

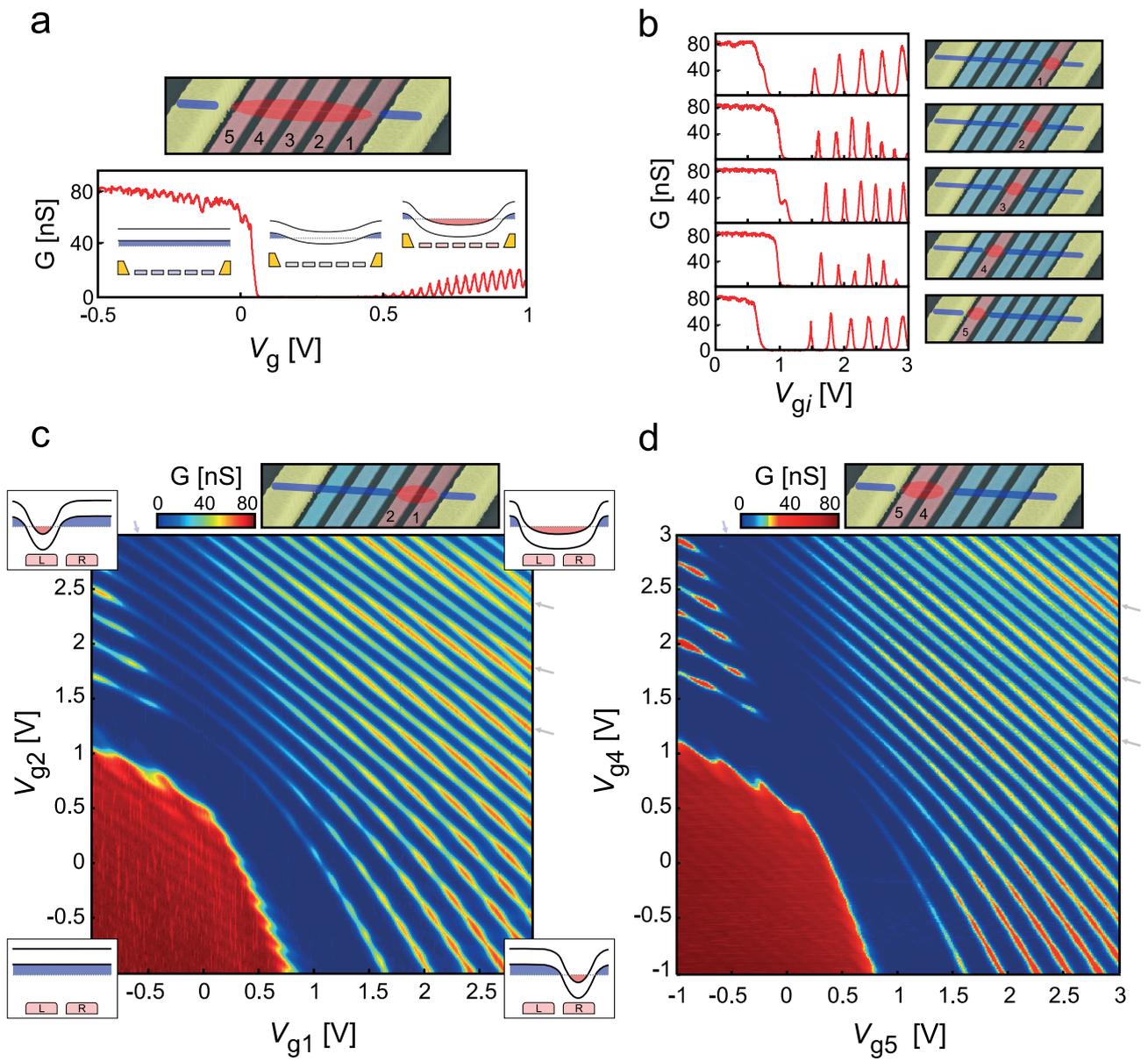

Figure 3

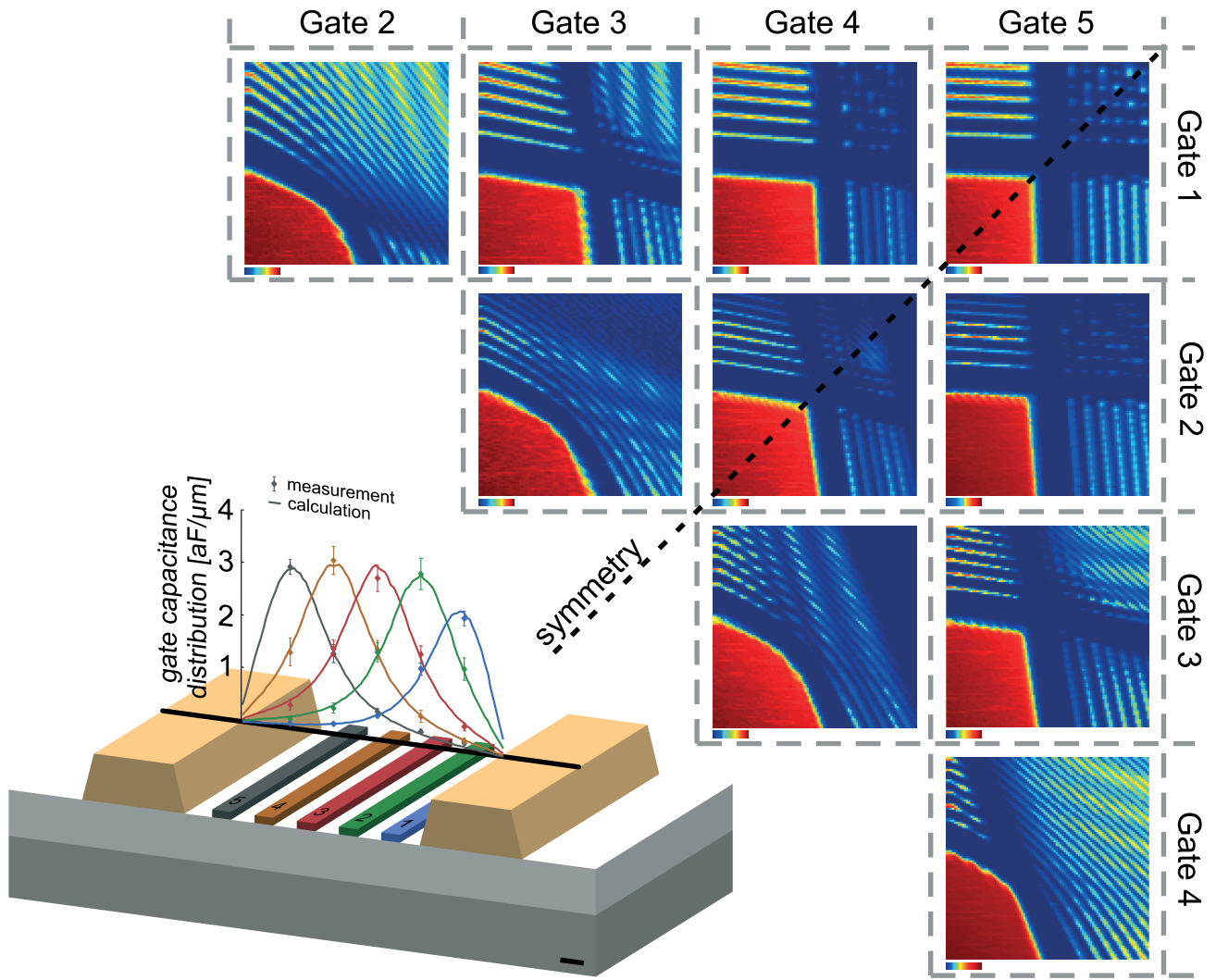

Figure 4

# SUPPLEMENTARY INFORMATION

**Realization of Pristine and Locally-Tunable One-Dimensional Electron Systems in Carbon Nanotubes**

J. Waissman, M. Honig, S. Pecker, A. Benyamini, A. Hamo and S. Ilani

S1. **Local cutting of the suspended NT**
S2. **Images of additional devices**
S3. **Details of the electrostatic calculation**
S4. **Quantitative estimate of electronic disorder strength**
S5. **Electrostatic analysis of the gate-gate conductance scans**
S6. **Designing electrostatic potentials with resolution determined by the gate pitch**
S7. **Transport data of a second device at dilution refrigerator temperatures**
S8. **Electrical functionality of the two-NT device as a coupled system-detector**

### S1.  Local cutting of the suspended NT

After the circuit is mated to a NT and *in-situ* transport measurements show the NT is clean and has the desired bandgap, we typically want to separate the device chip from the NT chip to allow transferring the prepared device for measurement in other setups. We achieve this separation by controlled cutting of the NT at well-defined positions, as explained in detail below.

The process is demonstrated in Figure S1a, which shows an SEM image of a seven-gated suspended NT device that has been selectively cut at two places (a magnified top-down view of these cuts is shown in the insets; the cuts are indicated by arrows). The relevant device segment is at the center, above the gates. To enable cutting at various locations we fabricate several contacts at each side of the device. The cutting process consists of applying a voltage between two adjacent contacts, which drives a large current through the short suspended NT segment between them. When the current passes a critical threshold the NT breaks at a single point, close to the center of the suspended segment. This cutting is believed to be due to Joule heating that leads to the highest temperature near the center of the suspended segment, which is farthest from the contacts that dissipate the heat.

The line traces in the insets of Fig S1a show the current-voltage characteristics measured during the cutting. As a function of the applied voltage the current grows monotonically, until reaching the critical current (15μA-30μA) and then dropping abruptly to zero, indicating that the segment between the contacts was cut. Measurement of transport through NT segments adjacent to the one that was cut before and after its cutting shows that they remain unaffected by this local process. Another technique that was found to efficiently cut the NT locally is the application of fast voltage pulses (typically ~10Volts/0.5μs) to one contact while its neighbor contact is grounded and all other contacts are floating. However, we generally prefer to do the cutting using the first approach (DC current) since it further allows us to distinguish between an individual single-wall NT vs. bundles or multi-wall NTs. For the latter, the cutting does not happen in a single step, but often exhibits multiple steps that correspond to the multiple tubes or multiple shells breaking one at a time. An example of such a two-stepped cut is shown in Fig. S1b. We observe the same pattern of steps when cutting the same NT at different junctions, demonstrating that these reflect the intrinsic properties of the tube and not of the junctions. If upon the first cut we observe any indication of a bundle or multiple shells we detach the circuit from the NT before performing the second cut and move to mate with a different tube. In general, we choose the growth parameters to yield sparse growth of suspended NTs, thereby avoiding bundles. However, if we find, using the above measurements, that a specific chip has a high density of NTs or indications of bundles we discard it.

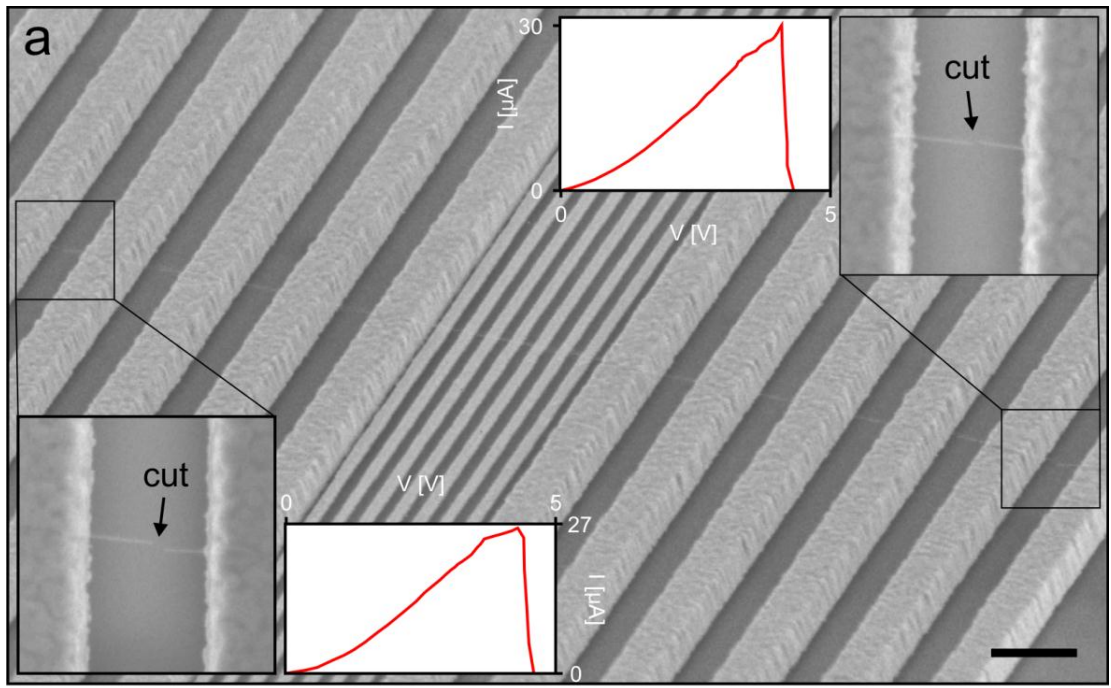

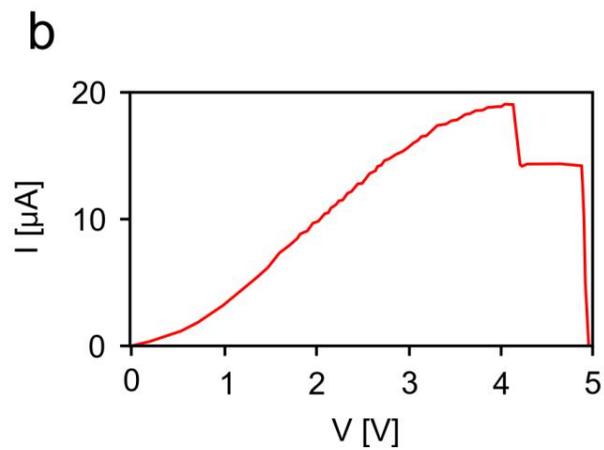

**Figure S1: The selective NT cutting process.** a) Main panel: Scanning electron micrograph of a seven-gated suspended NT device. Top-right and bottom-left insets: magnified top-down views of suspended segments that were cut using Joule heating with a flow of current (see text). The cut position is indicated by an arrow. Top right and bottom left traces: *I-V* curves of the burning process during voltage ramp-up. b) A cutting *I-V* curve during voltage ramp-up showing two abrupt current drops, attributed to the presence of a two-NT bundle or a double-walled NT.

## S2. Images of additional devices

In this section we show scanning electron microscope (SEM) images of three multi-gate devices formed using the mating technique, different than that shown in Figure 2c and d of the main text. Devices 2 and 3 have seven local gates, while Device 4 has five gates. These SEM images show that the NTs are anchored across the entire length of the contacts, are straight, and remain suspended above the gates with no slack, similar to the device shown in the main text. In fact, we find the vast majority of the devices made by our nano-assembly technique to have these ideal geometric features.

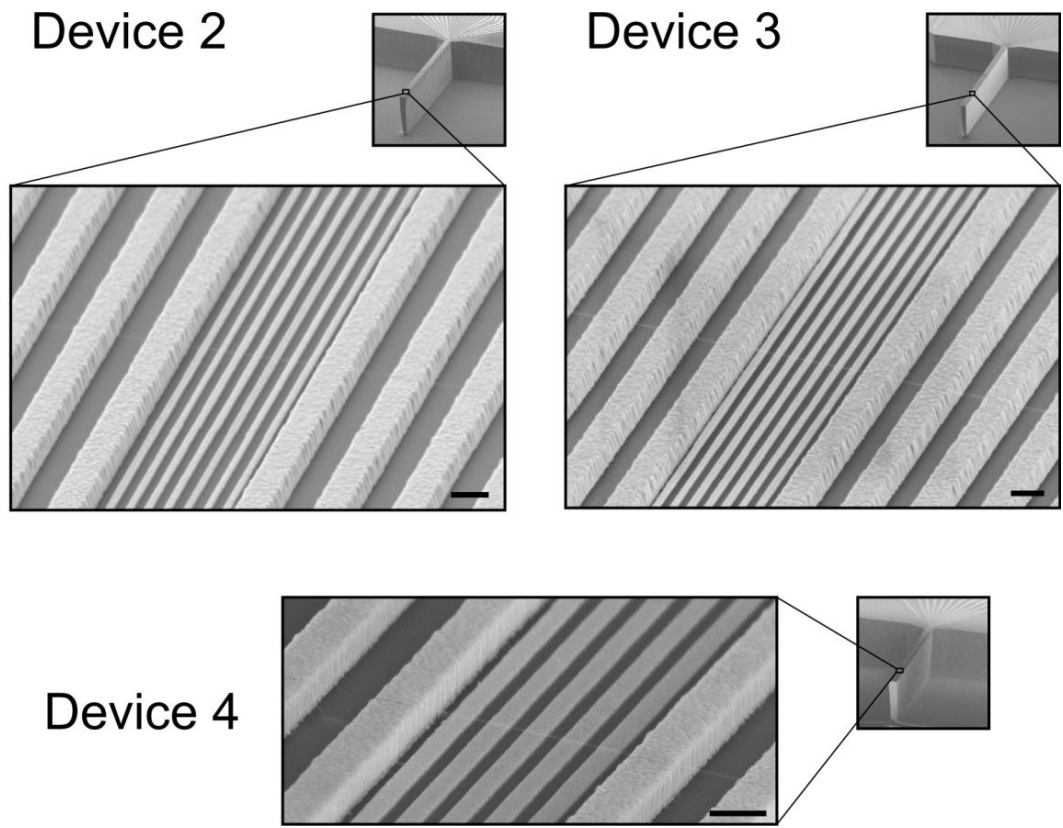

**Figure S2: Robustness of the mating technique: additional devices.** Scanning electron micrographs of circuits mated to suspended NTs (all scale bars = 300 nm). All devices underwent the cutting process, and all NTs remained intact and suspended over the gates, as shown. Devices 2 and 3 have seven gates of 65 nm width and 85 nm spacing. Device 4 has five gates of 120 nm width and 80 nm spacing.

## S3. Details of the electrostatic calculation

An important feature of the 1D multi-gated NT systems we present in this work is the ability to engineer any 1D potential profile along the NT length. Naively, the simplest way to create a potential $\phi(x)$, where $x$ is the position along the NT, is to have gates as close as possible to the NT such that their gating is local and their voltages directly determine $\phi(x)$. However, close proximity to metallic gates screens the interactions between electrons in the NT, thus destroying this salient feature. We therefore intentionally choose to distance the gates from the NT, the price being that gating becomes non-local, and a gate influences not only the NT segment above it but also segments above other gates. Knowing what gate voltages are required to produce a certain $\phi(x)$ thus necessitates quantitative knowledge of the non-local capacitive coupling to the NT. In Fig. 4 of the main text we show that we can extract this coupling directly from measurements with localized quantum dots, and that it corresponds quantitatively well to calculations, with no free parameters. In this section we explain in detail how the measurements of local quantum dots are translated to the spatial capacitance distribution of the gates, and the details of the finite element calculations.

The influence of a specific gate on the NT is fully captured by a capacitance distribution function $C_i(x) = en(x)/V_{gi}$, where $e$ is the electron charge, $i$ is the gate index and $n(x)$ is the charge distribution along the NT induced by a gate voltage $V_{gi}$, under the assumption that the NT is a perfect metallic conductor. We are interested in a discretized version of this function where the NT is partitioned to $N$ segments of equal length $l$ ($N$ is the number of gates) each segment being positioned above a corresponding gate. This partitioning reflects the "effective resolution" with which we can define the potentials with the gates. The capacitance of a gate $i$ to a segment $j$ is then given by $C_{ij} = \int_{segment\ j} C_i(x)\,dx \approx C_i(x_j) \cdot l$, where $x_j$ is the center coordinate of the segment. Experimentally we can extract a closely related quantity, by measuring the charge response of a quantum dot localized at position $j$ to the voltage on gate $i$. The latter amounts to $\tilde{C}_{ij} = \int_{dot\ j} C_i(x)\,dx \approx C_i(x_j) \cdot l_{dot\ j}$, where $l_{dot\ j}$ is the length of the dot formed at position $j$. Clearly, the $\tilde{C}_{ij}$'s depend on the shape of the quantum dots.

However, if we take only their ratios that measure the response of the *same* dot to two different gates, the details of the dot cancel out and we remain with the ratios of the quantities that we are seeking : $C_{ij}/C_{ii} = \tilde{C}_{ij}/\tilde{C}_{ii}$. These capacitive coupling ratios are directly extracted from the slope of the charging lines in the two-gate conductance scans shown in the multiple panels of figure 4 in the main text.

The ratios above give only relative capacitances, and thus do not provide the full information needed to determine all the absolute capacitance elements in the $C_{ij}$ matrix. To get the missing information we complement these data with measurements of the integrated capacitance of individual gates. To obtain these we form a large quantum dot extended over the entire suspended NT. We first measure the total capacitance of this dot to all five gates chained together. This quantity is directly extracted from the gate periodicity of the Coulomb oscillations in figure 3a of the main text. This capacitance gives the sum of all the $C_{ij}$ matrix elements: $C_{total} = \sum_{ij} C_{ij}$. Then we measure the relative contribution of each of the gates to this capacitance, giving the sum of one row in this matrix, $C_i/C_{total} = \sum_j C_{ij}/C_{total}$. We get the latter by comparing the width of a Coulomb peak of the large dot when only one gate is scanned vs. the width of this peak when all gates are scanned. Together, all these quantities give us the full capacitance matrix without any free parameters.

The electrostatic simulations are performed with the finite-element calculation package COMSOL. For this, we use the real device dimensions extracted from SEM images to model the geometry. This includes the trapezoidal cross-section of the contacts, arising due to gradual closing of the e-beam resist window during the thick evaporation. The electrode and substrate geometry that go into the calculation are shown in the bottom inset of Fig. 4 of the main text. We model the nanotube as a metallic cylinder resting on the contacts and suspended over the gates. To extract the capacitance distribution of gate $i$ we set the voltage on this gate to $V_{gi}$ while keeping the other gates, the contacts and the NT grounded. We then calculated the resulting charge distribution along the NT, $n(x)$, from which we get directly the capacitance distribution of this gate, $C_i(x) = en(x)/V_{gi}$. These capacitance distribution functions for the individual gates are shown in Fig. 4 of the main text, matching the experimental values with no free parameters.

## S4. Quantitative estimate of electronic disorder strength

As will be explained below, with the multiply-gated devices we can not only set the potential profile with the gates, but also estimate its magnitude in the absence of gating. The latter, which corresponds to the uncontrolled disorder potential fluctuations in our device, therefore provides an experimental estimate for the strength of disorder on length scales set by the gates' resolution. Our goal is thus to measure the uncontrolled potential modulations, $\delta\phi(x)$, that exist in the NT in the absence of gating.

If the NT had the same work function as the gate metal, it would be un-gated when all the gates are un-biased with respect to the NT ($V_{gi,i=1..5} = 0$). However, since generally these workfunctions are different, the absence of *electrochemical* bias on the gates ($V_{gi,i=1..5} = 0$) actually means that there is a non-zero *electrostatic* potential difference between the gates and the NT. This difference, termed the "contact potential", amounts to $\delta W_{gate-NT} = W_{gate} - W_{NT}$, and it gates the NT. To null this gating one must therefore apply a canceling electrochemical bias to the gates, $V_{gi,i=1..5} = -\delta W_{gate-NT}$. It is important to note that the contacts, which are by definition electrochemically shorted to the NT, produce a similar gating effect due to the difference between their workfunction and that of the NT, $\delta W_{contact-NT} = W_{contact} - W_{NT}$. This is the reason for the large hole doping of the NT segments that lie on top of the contacts, mentioned in the main text.

Combining the above understanding with finite element simulations, which were shown in the previous section to describe our system quantitatively well, we can determine the bare electrostatic potential produced along the NT for any combination of gate voltages:

$$\phi(x) = \sum_i \alpha_i(x) \cdot (V_{gi} + \delta W_{gate-NT}) + (\alpha_S(x) + \alpha_D(x)) \cdot \delta W_{contact-NT}, \qquad \text{Eq. S1}$$

Here $\alpha_i(x)$, $\alpha_S(x)$ and $\alpha_D(x)$ are unit-less functions, determined from the finite elements simulations, that give the potential along the NT per unit of voltage applied on gate $i$, the source and the drain respectively. The *a priori* unknown work function differences, $\delta W_{gate-NT}$ and $\delta W_{contact-NT}$, are the two free parameters of this equation that are determined from the experiments (see below).

To elucidate the relation between the potential profiles given by Eq. S1 and the measured transport we take as an example the top conductance trace from Fig. 3b in the main text that corresponds to the formation of a quantum dot above gate 1 (reproduced in Fig. S3a). In this scan $V_{g1}$ is swept while all the other gates are kept at fixed voltages $V_{g2..5} = -0.8V$. The calculated potential profiles, $\phi(x)$, that correspond to three gate voltages along this scan (circles in Fig S3a) are shown in Fig. S3b (work function differences are included; see below). In each of these plots the $\phi(x)$ in the suspended segment corresponds to the center of the NT bandgap as a function of position. Wherever it crosses from below to above the Fermi energy, $E_F$ (dashed horizontal line), the local occupation changes from holes to electrons.

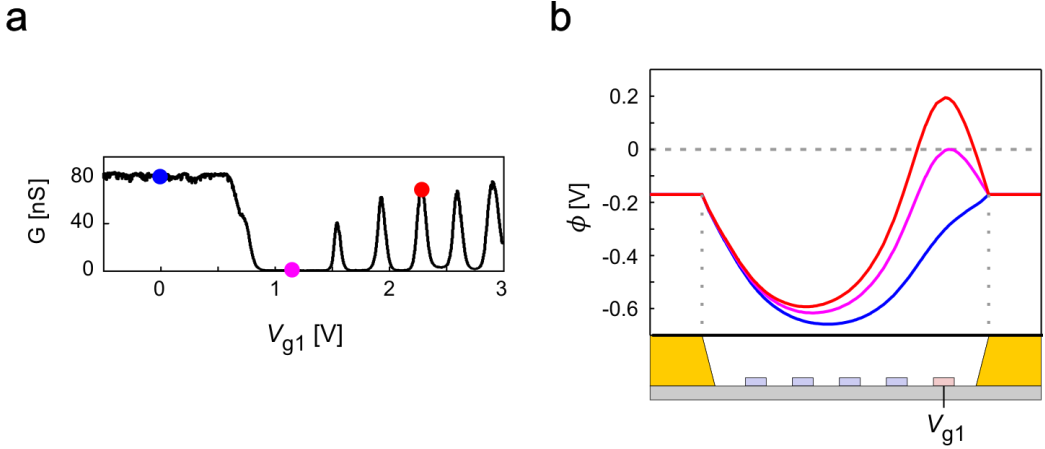

**Figure S3: Calculated bare electrostatic potential profiles along the NT for three different gating configurations.** a) Measured conductance trace for a dot formed above gate 1, equivalent to the top trace in figure 3b of the main text. In this measurement $V_{g1}$ is swept while the rest of the gates have a fixed potential $V_{g2..5} = -0.8V$ that dope the segment above them with holes. b) Three calculated potential profiles calculated using finite elements and Eq. S1, corresponding to three different transport regimes (the corresponding $V_{g1}$'s are shown in panel a as dots with similar colors). Workfunction differences are included (see text).

At low $V_{g1}$ the corresponding $\phi(x)$ (blue trace) is at all positions below $E_F$, implying that the NT is populated with holes over its entire length, thus forming a continuous "NT wire". At high $V_{g1}$, the corresponding $\phi(x)$ (red trace) exceeds $E_F$ above gate 1, crossing it at two points. At these points *p-n* junctions form, confining the electrons above gate 1 to a quantum dot. In between these regimes there is a $\phi(x)$ that exactly reaches $E_F$ above gate 1 (purple trace). For this potential the center of the NT bandgap is at $E_F$ above gate 1, forming a single long barrier above this gate. This point corresponds to the center of the non-conducting regime in the transport (purple point Fig S3a) where the conductance is maximally suppressed.

Looking at the corresponding transport traces obtained by scanning the other local gates (Fig S4a, reproduced from Fig 3b) we see that the center of the "gap" appears at very different gate voltages for the different gate positions. Fig S4b plots the gap-center gate-voltage as a function of the gate position, showing that this value changes by $\Delta V_g \approx 325 mV$ from the side gate to the center gate. This seemingly large potential modulation is in fact a direct result of the position dependence of the device electrostatics combined with the finite workfunction difference between the gates/contacts and the NT. Both these effects should be fully captured by Eq. S1. Thus, if this equation is accurate, in the absence of disorder we should be able to reproduce the position of these five gap centers, with just two parameters (the metal workfunctions). This is demonstrated in Fig. S4c, where we show the five $\phi(x)$'s that correspond to the gap centers in the five different conductance traces, calculated with $\delta W_{gate-NT} = -40 mV$ and $\delta W_{contact-NT} = -170 mV$. These work function values are consistent with published values for gold, palladium, and carbon NTs. We can clearly see that in all cases the potential reaches $E_F$ above the corresponding gate with an accuracy of $\delta \phi \approx \pm 5 mV$. These small fluctuations compared to those observed in Fig S4b ($\Delta V_g \approx 325 mV$) show that most of the effect is a consequence of the device electrostatics, and once it is known quantitatively it can be taken into account and nulled out. The remaining small fluctuations give us an upper bound of $\sim 5 mV$ on the magnitude of the bare potential disorder on the length scale set by the gate width. Using the lever-arm of the local gates ($\alpha \sim 0.3$) this is translated to $\sim 17 mV$ on a local gate, and by comparing this to the measured single-gate Coulomb blockade periodicity ($\Delta V_g \sim 315 mV$) necessary to introduce an electron charge above a gate we can obtain an upper bound on the local charge disorder of $\delta n \sim 5 \cdot 10^{-2} e$ on the gate length scale, a small fraction of a single electron charge. Alternatively, we can consider the induced fluctuations in the self-consistent disorder potential, where screening will reduce the bare disorder potential seen by electrons. By factoring in the ratio between the geometric capacitance ($\sim 4aF/\mu m$) and the quantum capacitance ($\sim 400 aF/\mu m$) of the nanotube, we estimate the self-consistent disorder potential at $\sim 50\mu V$. We note that this is a strict upper bound, since the potential fluctuations that we consider include all the errors in the measurements and calculations. The actual disorder is most likely significantly smaller. We also note that potential fluctuations on smaller length scales, which are too weak to form barriers for electron transport at the temperature of our measurements, would not be observed here.

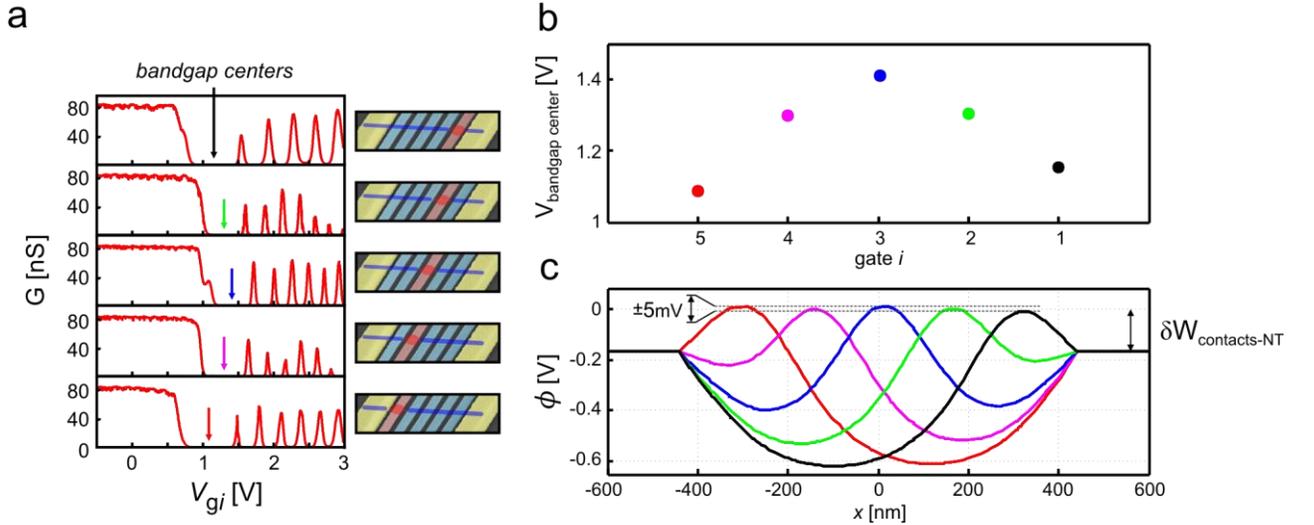

**Figure S4: Extracting an upper bound on the disorder potential from the measured transport.** a) The single-gate conductance scans reproduced from Fig 3b of the main text with the bandgap centers marked by colored arrows corresponding to the colored points in panel b and colored curves in panel c. b) The gap-center voltage as a function of the gate index, extracted from the graphs in panel a. c) The bare electrostatic potentials along the NT, $\phi(x)$, calculated using Eq. S1 for the gate voltage configurations that correspond to the positions indicated by the respective arrows in panel a, with $\delta W_{gate-NT} = -40mV$ and $\delta W_{contact-NT} = -170mV$. Although the gate voltage at the gap center varies between dots at different locations by as much as ~325mV, when we take into account the electrostatics of the device through Eq. S1 we see that within $\delta\phi = \pm 5mV$ all the potential profiles corresponding to the center of the gap are at the Fermi energy, giving an upper bound for the residual bare disorder potential fluctuations.

### S5. Electrostatic analysis of the gate-gate conductance scans

In this section, we analyze the features observed in the conductance map of Fig. 3c (and its symmetric partner in Fig. 3d). We use the electrostatic understanding established in section S3 and S4 above to determine the potential landscapes formed along the NT at the various gating configurations which correspond to different points in Fig. 3c and use these profiles to demonstrate the underlying sources of the observed features.

In figure S5a we reproduce Fig. 3c of the main text and highlight the secondary conductance features that are observed on top of the Coulomb blockade features described in the main text. The first set of features, apparent in this scan and its symmetric partner (Fig. 3d), are stripes of conductance modulation that are marked by dashed black lines. These stripes cross through the Coulomb charging lines and modulate their peak heights. The electronic configuration which corresponds to this region in the gate-gate diagram consists of an electron dot formed over gates 1 and 2 (red in device schematics) while above gates 3 to 5 the NT is populated with holes (blue in device schematics). The hole population in this segment is continuously connected to the holes above the left contact, forming a continuous "hole wire" that acts as a "NT lead" for the electronic quantum dot. We can confirm this picture by calculating the electrostatic potential induced by the gate voltages as shown in Fig.S4b, where the NT lead and the electron dot correspond to the potential well and hill above their respective gates. By calculating the potentials at two points along the modulation stripe (indicated in the figure by red and blue circles), we see that the potential well corresponding to the NT lead remains identical while only the electron potential hill has changed. This indicates that going along a stripe preserves the charge density in the NT lead, whereas going perpendicular to it changes this charge density. Thus, the origin of the observed striped modulations of the Coulomb peak heights is reproducible Fabry-Perot-like modulation of the conductance of the "NT lead" (the nature of these conductance modulations is discussed further in the last paragraph of this section).

We now proceed to confirm this picture with a calculation of the stripe slopes. Although gates 1 and 2 are far from the NT lead, they still gate it by an amount that can be quantitatively determined from the capacitance distribution functions which were measured and calculated (inset for Fig 4 in the main text). The relative capacitance of the two gates to the NT lead amounts to the ratio of the areas under the capacitance distribution curves of these two gates integrated over the length of the NT lead, as shown in Fig. S5c. The edge of the hole occupation is determined from the point where the electrostatic potential $\phi(x)$ crosses zero, since this is where the *p-n* junction barrier will be centered. In the discretized version these capacitances are given by the elements of the capacitance matrix, $C_1^{p-lead} = C_{13} + C_{14} + C_{15}$ and $C_2^{p-lead} = C_{23} + C_{24} + C_{25}$, all of which we measure directly. The dashed black lines in figure S5a are drawn with a slope $C_1^{p-lead}/C_2^{p-lead}$ taken from these measurements, showing a good fit to the observed conductance modulation slopes.

Another clear feature observed in the two, mirror-symmetric, gate-gate scans is a band of suppressed conductance (dashed white lines, Fig S5a). We calculate the electrostatic potential in the middle of this suppressed band along the same line of fixed NT lead gating, shown in green in Fig. S5b. From the electrostatic calculations we can identify that this feature corresponds to having the electron dot confined over only a single gate (gate 2) and having the NT bandgap pinned between gate 1 and the right contact. In this case, the right tunnel barrier of the electron dot is a *p-n* junction formed above an edge gate (gate 1 or gate 5), which has a longer depletion length than when formed above a center gate because the nearby contact is grounded (as opposed to the negatively-biased gates). The underlying origin of the longer barrier is the shallower slope of $\phi(x)$ where intersects zero on the right. The existence of this extended edge barrier explains the observed conductance suppression. The slope of the dashed white line in the conductance map that follows the middle of the observed suppression band is $\frac{\Delta V_2}{\Delta V_1} \approx 2.5$, corresponding reasonably well to that calculated with the capacitance distributions, $C_1^{p-n\,barrier}/C_2^{p-n\,barrier} \approx C_{11}/C_{12} = 2.3$. We note here that the actual length of the *p-n* junction will depend on electrostatic and quantum effects, and determining it requires a full solution of the Schrodinger-Poisson self-consistent equations. However, we are interested only in the positions of features in the voltage-voltage plane, which depend on the position of the *p-n* junction (and not its width), and this is captured well by our analysis.

Continuing to the other side of the suppression feature, the calculated electrostatic potential (gray line, Fig. S5b) shows that the quantum dot remains over gate 2, but above gate 1 there is now a hole population, showing that the right "NT lead" has extended over gate 1. As a result, the right *p-n* junction is now shorter and the conductance is higher, comparable to that on the other side of the suppression feature.

Finally, we would like to comment on the nature of the hole-doped "NT leads" to the electronic dot. In the main text we mentioned that when the NT is populated entirely with holes it behaves like a "NT wire". In this regime we measure only weak gate modulation of the conductance, which phenomenologically resembles the measurements of NTs in the Fabry-Perot regime[1]. In that regime the barriers between the NT and the contacts are highly transparent and the average conductance is comparable to $4e^2/h$. In our case, on the other hand, the measured conductance is significantly smaller than the quantum conductance and thus one expects the NT to behave as a quantum dot rather than a Fabry-Perot cavity. This dot, however, is unusual since its charging energy is strongly suppressed. This suppression results from the fact that the NT sits directly over the contacts and thus has an extremely large capacitance to them. At such short distances the geometrical capacitance to the contact, $C_{source}^{geometrical}$, is much larger than the quantum capacitance of the NT segment above it, $C_{source}^{quantum}$, so that the latter dominates the total source capacitance, which in our case is $C_{source} = \left[ {C_{source}^{geometrical}}^{-1} + {C_{source}^{quantum}}^{-1} \right]^{-1} \approx 100aF$. The resulting charging energy of the NT leads, $U = e^2/(C_{source} + C_{drain} + \sum C_{gates})$, being dominated by $C_{source}$, thus roughly equals the level spacing of the NT above the contacts. In this respect, the system is similar to the Fabry-Perot cavity. The large suppression of the charging energy as compared to the quantum dots on the suspended part of the NT explains why at $T = 4K$ the hole-doped NT behaves similarly to a Fabry-Perot cavity and shows weak Coulomb oscillations. This is the regime of the "NT wire" in our measurements, where the charging energy, ~1mV, is small enough that the measurement temperature results in only weak gate-dependence of the conductance, and the series resistance to the metallic contacts nonetheless gives a small overall conductance. For a given overlap length of NT and contacts, when the temperature is low enough the leads would eventually show Coulomb blockade physics. For the device geometry shown in Fig. 2c such Coulomb physics of the leads is indeed seen at dilution temperatures (see section S6). However, by making the overlap with the contacts long enough, and hence suppressing further the charging energy, it should be possible to make the leads behave as "wires" down to the lowest temperatures in our measurements.

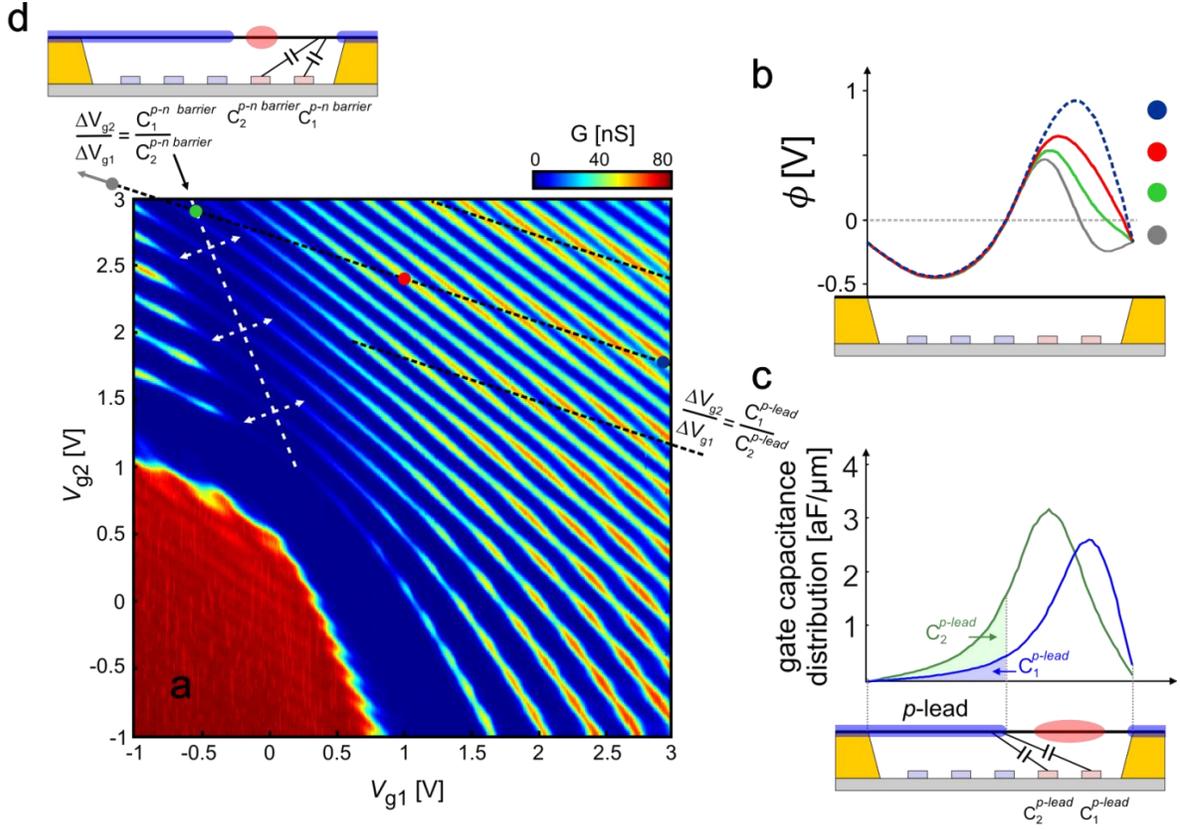

**Figure S5: The electrostatic origin of the observed features in the two-gate conductance scans.** a) The main panel (duplicating Fig. 3c of the main text) shows conductance, $G$, as a function of $V_{g1}$ and $V_{g2}$ while $V_{g3-5}$ = -0.8 V. The dashed black lines and dashed white lines correspond to the secondary conductance features observed in this scan. These lines are drawn with slopes taken directly from the measured capacitances (see text and panels below). b) The electrostatic potential along the NT length, $\phi(x)$, calculated for the points indicated in panel a, with circle colors corresponding to line colors. The contact and gate work function differences are included as described in the text. The device schematic is to scale in both dimensions. c) Schematics of the electron (red) and hole (blue) population along the NT corresponding to the blue and red points in panel a. The top traces show the capacitance distributions of gates 1 and 2. The colored areas under the curves give the capacitances between these gates and the left NT hole "lead" to the dot. The ratio between these capacitance gives the slope of the dashed black lines in panel a d) Schematics of the electron and hole population along the NT corresponding to the green point in panel a. Here a dot forms only above gate 2 and the NT is in the gap over gate one. This leads to an extended right barrier that yields the suppression along the dashed white line in panel a. Its slope in the voltage-voltage plane is determined by the ration of the capacitances of the barrier region to gates 1 and 2.

## S6. Designing electrostatic potentials with resolution determined by the gate pitch

In this section we demonstrate that with knowledge of the electrostatic coupling of the gates to the NT (as demonstrated in section S4 above), we can design potential profiles along the NT with a spatial resolution given by the gate pitch and not smeared by the separation between the NT and the gates. As explained in the main text, by distancing the nanotube from the gates, we preserve electron-electron interactions. But at the same time we also spatially smear the effect of individual gates on the NT. While a close gate controls the potential in the NT along a length comparable to its width, a distant gate affects a longer section amounting to the convolution of the gate width and its distance to the NT. This distance therefore reduces the effective resolution with which we can design electrostatic potentials. However, by using our knowledge of the non-local gate coupling, we can deconvolve this spatial smearing and define potential features whose sharpness is determined by the gate pitch alone. Such deconvolution works as long as the NT is not too far from the gates compared to the gate separation.

To define the potential (or the charge) on the NT with gate pitch resolution means that if we partition the NT into $N$ segments of equal length, where $N$ is the number of gates, we can define the potential (or charge) in each one of these segments independently. However, the charge on the $i^{th}$ segment in the NT, $q_i$, due to a voltage on gate $j$, $V_{gj}$, is given by $q_i = C_{ij} \cdot V_{gj}$, where $C_{ij}$ is the capacitance coupling matrix element. Thus, the above equation shows that a gate does not only affect the local segment above it but also neighboring segments, reducing the effective resolution. To define the charge on each segment independently, we instead *invert* the equation, $V_{gj} = C_{ji}^{-1} \cdot q_i$, to obtain the linear combination of gate voltages $V_{gj}$ that is needed to control the charge in only a *single* segment of the NT, $q_i$. This inversion amounts to a discretized deconvolution of the capacitive smearing.

Figure S6 illustrates how this deconvolution works for the dimensions of our devices (a NT-gate distance of 130nm and a gate pitch of 150nm), using a calculation of the potential at the NT, $\phi(x)$, with Eq.S1, Applying a voltage on a single gate (Fig. S6a) leads to a potential along the NT spread out over ~325nm (Fig. S6b), roughly the sum of the gate pitch and the NT-gate distance. On the other hand, if we use instead the linear combination of gate voltages found using the inverse capacitance matrix (Fig. S6c), we obtain a potential that is ~160nm wide (Fig.S6d), comparable to the gate pitch. We note that the deconvolution becomes exponentially harder when the NT-gate distance becomes much larger than the gate pitch, $\eta \equiv \frac{d_{NT-gate}}{d_{gate-gate}} \gg 1$, since in this regime the voltages necessary for producing the desired potentials increase exponentially in $\eta$, and any experimental error in determining $C_{ij}$ is exponentially amplified. However, as long as $\eta$ is not very large, as in our experiments where $\eta \sim 1$, the above deconvolution procedure works well.

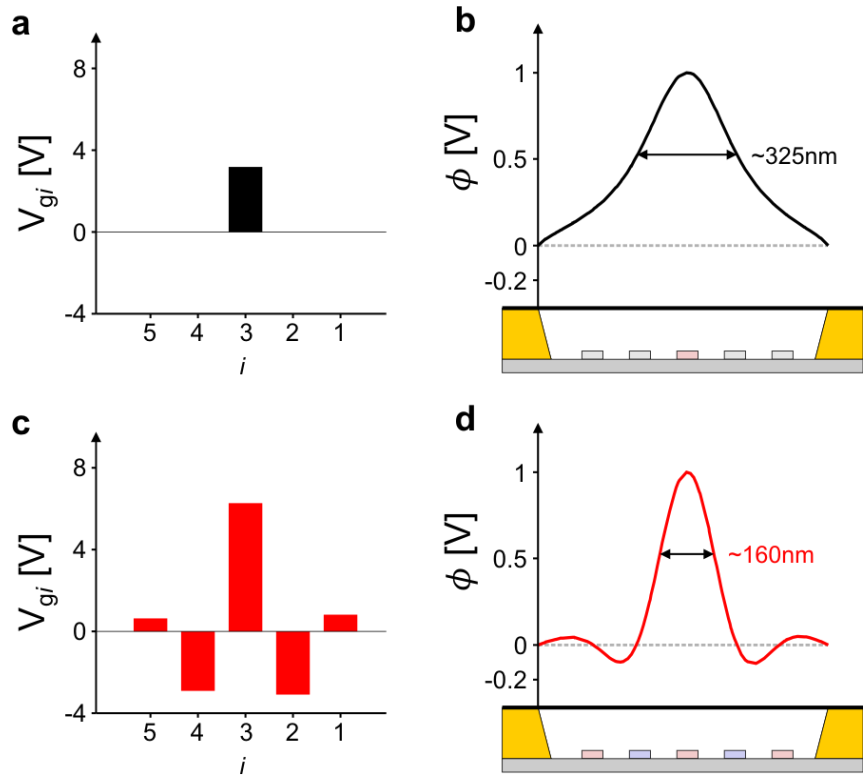

**Figure S6: Recovering the gate resolution with potential design.** a) The voltages applied on the gates and b) the corresponding calculated potential at the NT, $\phi(x)$, for the dimensions of our device (a NT-gate distance of 130nm and a gate pitch of 150nm). In this example a voltage is applied only on a single gate and the width of the potential feature along the NT is roughly the sum of the gate pitch and NT-gate distance. c) Linear combination of voltages for producing localized potential above the central gate, obtained by inverting the measured capacitance matrix. d) The corresponding calculated potential along the NT showing that the non-locality of the gate coupling can be effectively deconvolved.

## S7. Transport data of a second device at dilution refrigerator temperatures

The device shown in the main text showed no observable indications of disorder at the energy scale of the measurement ($T = 4K$). A natural question is whether at lower temperatures, smaller disorder scales would become observable, and we address this with measurements at dilution refrigerator temperatures. Figure S6 shows the conductance of a five-gated device, different than the one shown in the main text, measured in a dilution refrigerator with a base temperature of $T = 7mK$ (extracted electron temperature is $T \approx 80mK$). At these temperatures, the device is expected to be sensitive to smaller magnitude disorder; we show in the following that our observations on device cleanliness hold to these low temperatures. For this experiment, the two right gates are biased together along the horizontal axis, $V_{g1} = V_{g2} = V_R$, the two left gates are biased together along the vertical axis, $V_{g4} = V_{g5} = V_L$, and the center gate is biased with the average voltage, $V_{g3} = (V_L + V_R)/2$.

Overall, this device demonstrates almost perfectly clean behavior. In the upper right (lower left) corners, we observe the creation of a five-gate electron (hole) dot (see corresponding schematics) with a single Coulomb oscillation periodicity. In the lower right and upper left corners, *p-n* junctions are formed at the center of the suspended NT, leading to the creation of a hole-electron and electron-hole double quantum dots respectively (see corresponding schematics). Notably, every vertical charging line in the lower right corner, corresponding to an electron localized on the right side of the device, evolves smoothly into a horizontal charging line in the upper left corner, corresponding to an electron localized on the left side. As explained in the main text this smooth evolution shows that individual electrons are smoothly shuttled from the right to the left side of the device without apparent effects of disorder. An almost perfectly symmetric behavior is observed for the hole charging lines that evolve smoothly from vertical in the top left corner to horizontal in the bottom right corner. One deviation from the perfect behavior is observed for the first hole line, which is vertical even on the bottom part of the gate-gate scan and does not bend like the others. This means that this hole gets stuck on the right side and is not shuttled to the left side by the gates. By checking the relative coupling of this feature to the individual gates (not shown) we see that it almost exclusively gated by gate 5 and none of the others, demonstrating that this hole is localized between gate 5 and the contact. Such behavior could be due to a highly localized potential dip near the contact that binds only one carrier. Importantly, all the holes after this first localized one show the normal extended behavior and possess nearly perfect symmetry to the electrons.

Looking carefully on the data for the first few electrons and holes (excluding the first localized hole), we can observe small wiggles of the charging lines. These wiggles are an order of magnitude smaller than those observed in the best ultra-clean double-dot devices made to date[2], demonstrating that the underlying disorder potential in our devices is much smaller. As was clearly demonstrated[2], a potential hump or dip act differently on electron and holes, leading to different charging line structures for the two carrier types. The fact that we observe very similar wiggles for electrons and holes therefore emphasizes that disorder on the length scale of the gate spacing is probably not the mechanism leading to the observed wiggles. Instead, the effect must operate the same way on electrons and holes. One candidate is the attraction of the carriers in the NT to their image charges formed at the metallic leads, which leads to attractive potentials at the suspended NT edges for both electrons and holes, thereby leading to a double-dot-like effective potential which could explain the small wiggles. Another plausible mechanism is strong interactions between the carriers, which are predicted to lead to real-space separation of charge carriers and a similar modulation of the charging lines.

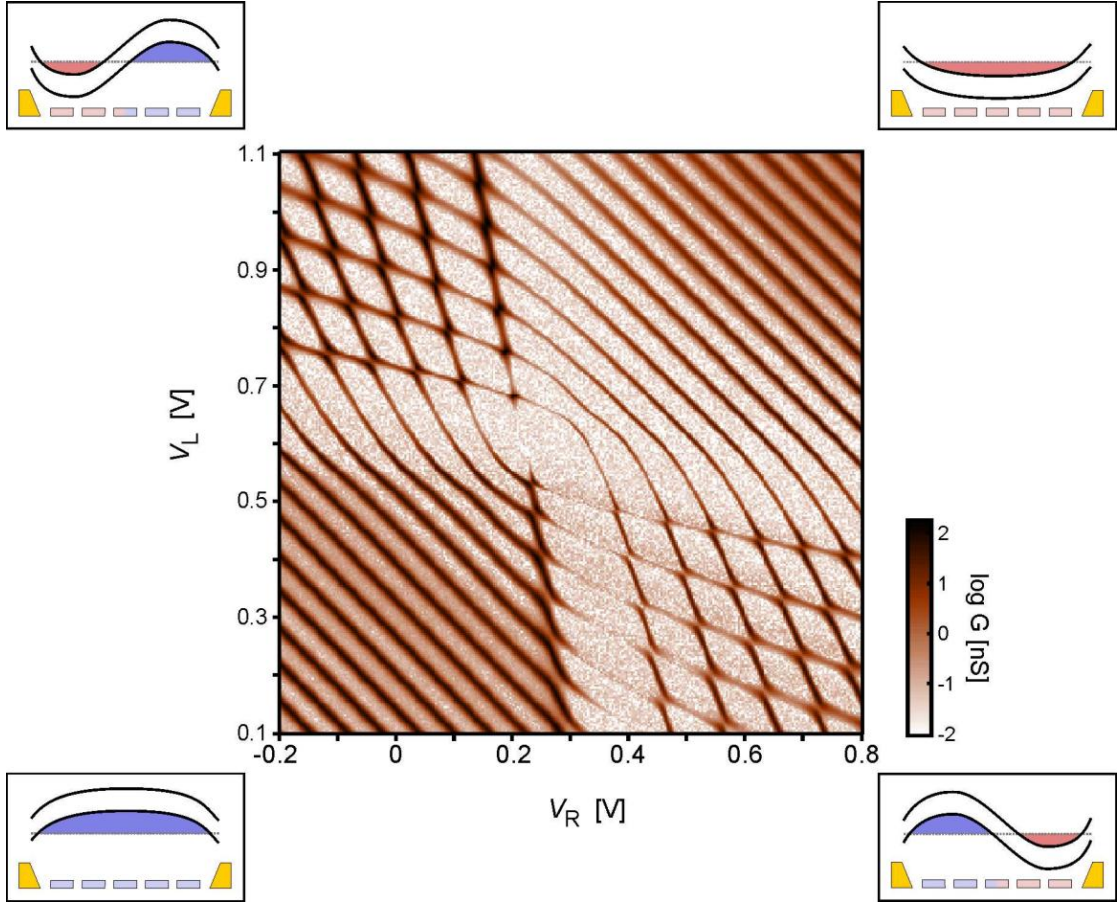

**Figure S7:** A **second five-gated small bandgap NT device measured at dilution refrigerator temperatures.** The conductance, *G*, on a logarithmic scale, measured as a function of right and left gate voltages, $V_R$ and $V_L$. The right gate voltage is applied on the two right gates, $V_{g1} = V_{g2} = V_R$, the left gate voltage is applied on the two left gates, $V_{g4} = V_{g5} = V_L$ and on the center gate we apply the averaged value $V_{g3} = (V_R + V_L)/2$. The insets show schematic band diagrams corresponding to the four quadrants of the measurement: bottom left - a hole dot over all gates; top right - an electron dot above all gates; bottom right - electron/hole double dot; top left - hole/electron double dot.

**S8. Electrical functionality of the two-NT device as a coupled system-detector**

In the main text, we demonstrated the ability to place two nanotubes in the same device at a controlled separation (Fig.2d). This device geometry enables new device functionality: the ability to use one nanotube as a quantum-dot detector to electrically sense the second tube. This circuit implementation, new to suspended NTs, has a large number of potential applications, including: charge detection in ultra-clean 1D systems, charge measurement in quantum information implementations, and measurements of mechanical motion of NT nano-mechanical resonators. A few works in the past have incorporated local detectors into nanotube circuits[3,4], however, so far these devices have been limited to nanotubes lying on a substrate, and local detection of an ultra-clean suspended nanotube has remained a challenging goal. Here, we demonstrate the possibility to perform such detection in a suspended device using the simplest example, in which we use a quantum dot on one tube to electrically sense the mechanical oscillations of the second tube. This approach is reminiscent to the single electron transistor (SET) motion detection performed on bulk silicon nano-beams[5], but here it is shown for the first time in the important context of multi-gated suspended nanotube. We note that due to the difficulty to make good SETs or quantum-dot detectors close to a NT mechanical resonator, to date, all studies of NT mechanical motion have used the gate-dependent transport through the moving NT itself to detect the motion. While this detection scheme has been very fruitful in past experiments[6–11], it puts important constraints on which measurements can be performed. Since the detection needs the transport through the resonator to be gate-dependent, it cannot be used, for example, when the transport is blockaded (e.g. within a Coulomb blockade valley) or conversely when it is in a 'metallic wire' regime having no gate dependence. Using an external quantum-dot detector to detect the movement, as we demonstrate below, decouples the mechanical and the detection components, alleviating the above constraints.

The measurement circuit is shown in figure S8. The left segment of the bottom NT, contacting the left and middle contacts and suspended above a single gate is used as the quantum-dot detector. A DC voltage on this local gate, $V_g^{det}$, creates a dot of electrons (marked red in the figure) and brings its Coulomb blockade transport to a point that is

20**S8. Electrical functionality of the two-NT device as a coupled system-detector**

In the main text, we demonstrated the ability to place two nanotubes in the same device at a controlled separation (Fig.2d). This device geometry enables new device functionality: the ability to use one nanotube as a quantum-dot detector to electrically sense the second tube. This circuit implementation, new to suspended NTs, has a large number of potential applications, including: charge detection in ultra-clean 1D systems, charge measurement in quantum information implementations, and measurements of mechanical motion of NT nano-mechanical resonators. A few works in the past have incorporated local detectors into nanotube circuits[3,4], however, so far these devices have been limited to nanotubes lying on a substrate, and local detection of an ultra-clean suspended nanotube has remained a challenging goal. Here, we demonstrate the possibility to perform such detection in a suspended device using the simplest example, in which we use a quantum dot on one tube to electrically sense the mechanical oscillations of the second tube. This approach is reminiscent to the single electron transistor (SET) motion detection performed on bulk silicon nano-beams[5], but here it is shown for the first time in the important context of multi-gated suspended nanotube. We note that due to the difficulty to make good SETs or quantum-dot detectors close to a NT mechanical resonator, to date, all studies of NT mechanical motion have used the gate-dependent transport through the moving NT itself to detect the motion. While this detection scheme has been very fruitful in past experiments[6–11], it puts important constraints on which measurements can be performed. Since the detection needs the transport through the resonator to be gate-dependent, it cannot be used, for example, when the transport is blockaded (e.g. within a Coulomb blockade valley) or conversely when it is in a 'metallic wire' regime having no gate dependence. Using an external quantum-dot detector to detect the movement, as we demonstrate below, decouples the mechanical and the detection components, alleviating the above constraints.

The measurement circuit is shown in figure S8. The left segment of the bottom NT, contacting the left and middle contacts and suspended above a single gate is used as the quantum-dot detector. A DC voltage on this local gate, $V_g^{det}$, creates a dot of electrons (marked red in the figure) and brings its Coulomb blockade transport to a point that is sensitive to external gating. The top, longer NT forms the mechanical resonator, whose mechanical vibrations are measured using a simple adaptation of a standard mixing technique[6]: A frequency-modulated (FM) radiofrequency (RF) signal with a carrier frequency $f$ is applied on the source contact of the detector quantum dot. This contact, being only $85nm$ away from the long NT, couples to it capacitively, actuating its mechanical motion when $f$ is resonant with one of its mechanical modes. This motion, in turn, produces a fluctuating gate potential on the quantum-dot detector. The size of the oscillating gate potential produced by the mechanical motion is $\delta V_{elec} = \delta z_{mech} \frac{1}{C}\frac{dC}{dz} V_{res-det}$, where $\delta z_{mech}$ is amplitude of the mechanical vibration, $C$ and $dC/dz$ are the capacitance between the resonator and detector NTs and its derivative with respect to their mutual distance, and $V_{res-det}$ is the (externally-controlled) bias between the resonator and detector circuits[6]. This oscillating gate signal is mixed down with the FM signal transmitted directly to the quantum dot by its source contact, through the finite transconductance of the dot's transport, $d^2I/dV_g dV_{sd}$, producing a low-frequency mixing current detected at its drain (middle contact) using a lock-in amplifier operating at the FM modulation frequency.

Figure S8b shows the out-of-phase component of the mixing current measured as a function of $V_{res-det}$ and $f$, at $T = 4K$. In addition to controlling the amplitude of the detected signal, the voltage difference $V_{res-det}$ applies also a mechanical force that tensions the long NT resonator. This tensioning leads to an increase of the frequencies of its mechanical modes. Fig. S8b shows one such mechanical resonance, visible as a peak in the mixing current, exhibiting a parabolic dependence of its frequency on $V_{res-det}$.

It is important to note that the above measurement was performed when the resonator NT was electrically tuned to be in the 'hole wire' regime, in which its transport is practically gate-independent. The absence of gate dependence would have not allowed the measurement of the mechanical resonance in the conventional way that uses transport through the resonator itself. However, such measurement becomes possible here by using the separate quantum-dot detector that can be tuned to a gate sensitive point, independent of the state of the resonator.

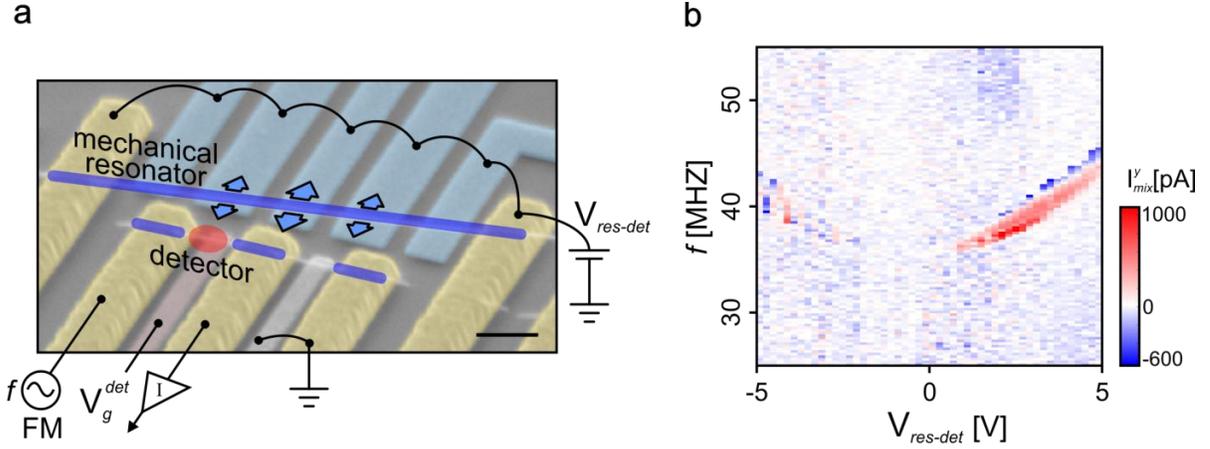

**Figure S8: Electrical detection of a NT mechanical resonator motion using a second NT quantum dot detector.** a) Measurement circuit, overlaid over the SEM image of the two-NT device from Fig. 2d of the main text, that was used in these measurements. Yellow: contacts. Blue/red/white: gates. The NTs are colored in red (blue) to reflect regions that are electron (hole) doped. The red blob represents a quantum dot formed on the left side of the bottom NT, suspended between two contacts, using a voltage on the gate beneath it, $V_g^{det.}$. The second NT, doped with holes over its entire length, is driven into motion (indicated by blue arrows) by a frequency-modulated (FM) radiofrequency (RF) signal with a carrier frequency $f$ applied on the left contact of the dot, that is capacitively coupled to it. The quantum dot acts as a non-linear mixing element, which mixes the RF signal on its source contact and the RF gating signal produced by the oscillating nanotube resonator, to a measurable low frequency signal measured at the drain of the dot (middle bottom contact). The right contact and gate are grounded during the measurement. The contacts and gates of the resonator tube are all kept at the same potential, $V_{res-det}$, which we control. Scale bar 300nm. b) The out-of-phase component of the mixing current, $I_{mix}^y$ (colormap), measured as a function of the bias between the resonator and detector circuits, $V_{res-det}$, and the FM carrier frequency $f$. The signal vanishes everywhere except at a mechanical resonance of the resonator, whose frequency increases with increasing $V_{res-det}$ due to electrostatic tensioning of the NT resonator.

# References

1. Liang, W. *et al.* Fabry - Perot interference in a nanotube electron waveguide. *Nature* **411**, 665–9 (2001).

2. Steele, G. a, Gotz, G. & Kouwenhoven, L. P. Tunable few-electron double quantum dots and Klein tunnelling in ultraclean carbon nanotubes. *Nature nanotechnology* **4**, 363–7 (2009).

3. Gotz, G., Steele, G. a, Vos, W.-J. & Kouwenhoven, L. P. Real time electron tunneling and pulse spectroscopy in carbon nanotube quantum dots. *Nano letters* **8**, 4039–42 (2008).

4. Churchill, H. O. H. *et al.* Relaxation and Dephasing in a Two-Electron C13 Nanotube Double Quantum Dot. *Physical Review Letters* **102**, 2–5 (2009).

5. Naik, a *et al.* Cooling a nanomechanical resonator with quantum back-action. *Nature* **443**, 193–6 (2006).

6. Sazonova, V. *et al.* A tunable carbon nanotube electromechanical oscillator. *Nature* **431**, 284–7 (2004).

7. Witkamp, B., Poot, M. & Van der Zant, H. S. J. Bending-mode vibration of a suspended nanotube resonator. *Nano letters* **6**, 2904–8 (2006).

8. Peng, H., Chang, C., Aloni, S., Yuzvinsky, T. & Zettl, a. Ultrahigh Frequency Nanotube Resonators. *Physical Review Letters* **97**, 087203 (2006).

9. Lassagne, B., Aguasca, A. & Bachtold, A. Ultrasensitive Mass Sensing with a Nanotube Electromechanical Resonator. *Nano letters* **2**, 4–7 (2008).

10. Wang, Z. *et al.* Phase transitions of adsorbed atoms on the surface of a carbon nanotube. *Science (New York, N.Y.)* **327**, 552–5 (2010).

11. Wu, C. C., Liu, C. H. & Zhong, Z. One-step direct transfer of pristine single-walled carbon nanotubes for functional nanoelectronics. *Nano letters* **10**, 1032–6 (2010).